\documentstyle[12pt,epsfig,a4]{article} 
\newcommand{\vs}{\vspace{-0.25cm}}

\newcommand{\vlk}{$V_{{\rm low}-k}$ }
\newcommand{\vlkn}{$V_{{\rm low}-k}$}
\newcommand{\be}{\begin{equation}}
\newcommand{\ee}{\end{equation}}

\begin{document} 

\begin{center}
\large{\bf Chiral three-nucleon interaction and the $^{14}$C dating  
beta decay}\footnote{Work supported in part by BMBF, GSI and by the DFG
cluster of excellence: Origin and Structure of the Universe.}

\bigskip 

J.\ W.\ Holt, N.\ Kaiser, and W.\ Weise\\

\bigskip

\small{Physik Department, Technische Universit\"{a}t M\"{u}nchen,\\
    D-85747 Garching, Germany}

\end{center}

\bigskip

\begin{abstract}
We present a shell model calculation for the beta decay of $^{14}$C to the
$^{14}$N ground-state, treating the relevant nuclear states as two $0p$-holes
in an $^{16}$O core. Employing the universal low-momentum nucleon-nucleon 
potential $V_{\rm low-k}$ only, one finds that the Gamow-Teller matrix element
is too large to describe the known (very long) lifetime of $^{14}$C. As a novel
approach to the problem, we invoke the chiral three-nucleon force (3NF) at 
leading order and derive from it a density-dependent in-medium NN interaction.
Including this effective in-medium NN interaction, the Gamow-Teller 
matrix element vanishes for a nuclear density close to that of saturated
nuclear matter, $\rho_0 = 0.16\,$fm$^{-3}$. The genuine short-range part of
the three-nucleon interaction plays a particularly important role in this
context, since the medium modifications to the pion propagator and 
pion-nucleon vertex (due to the long-range 3NF) tend to cancel out in the 
relevant observable. We discuss also uncertainties related to the off-shell 
extrapolation of the in-medium NN interaction. Using the off-shell behavior of 
$V_{\rm low-k}$ as a guide, we find that these uncertainties are rather small.
\end{abstract}

\bigskip
{\small PACS: 21.30.Fe, 21.60.Cs, 23.40.-s}

\section{Introduction}
The anomalously long beta decay lifetime of $^{14}$C, which makes possible the 
radiocarbon dating method, has long been a challenge to nuclear structure 
theory. The transition from the $J_i^\pi = 0^+$, $T_i = 1$ ground state of 
$^{14}$C to the $J_f^\pi = 1^+$, $T_f = 0$ ground state of $^{14}$N is of the 
allowed Gamow-Teller type, yet the known lifetime of $\sim 5730 \pm 30$ years is 
nearly six orders of magnitude longer than would be expected from typical 
allowed transitions in $p$-shell nuclei \cite{ajzen,chou}. The associated 
Gamow-Teller (GT) transition matrix element for the $^{14}$C decay must 
therefore be accidently small, on the order of $2 \times 10^{-3}$, which makes 
this transition a sensitive test for both nuclear interaction models and 
nuclear many-body methods.

The earliest studies to give insight into this problem were based on 
phenomenological models of the residual nuclear interaction. If only central 
and spin-orbit forces are included, it can be shown \cite{inglis} that it is 
impossible to achieve a vanishing GT matrix element when the model space is 
restricted to $0p^{-2}$ configurations consisting of two $0p$ holes in an 
$^{16}$O core. Later, Jancovici and Talmi discovered \cite{talmi} that this 
problem can be overcome with the addition of a tensor force component in the 
residual interaction. When realistic nuclear forces based on meson 
exchange were developed in the 1960's and applied to nuclear many-body problems 
through the $G$-matrix formalism, Zamick found \cite{zamick66} the decay rate 
to be very sensitive to the $0p_{1/2}-0p_{3/2}$ splitting, with overall 
unsatisfactory results when the experimental value of 6.3 MeV is used. A more 
recent calculation \cite{aroua} of the $^{14}$C beta decay has been performed 
in a large-basis no-core shell model study performed with the Argonne V8$^\prime$ 
interaction \cite{argv8}. Although it was shown that the inclusion of additional 
configurations up to $6\hbar \omega$ lead to a suppression of the GT matrix 
element, the results were not yet converged at this order.

Very recently, it has been suggested \cite{holt} that the $^{14}$C beta decay 
transition matrix element should be particularly sensitive to the density-dependence 
of the nuclear interaction. The study in ref.\ \cite{holt} used a medium-dependent 
one-boson-exchange nuclear interaction modeled with Brown-Rho scaling 
\cite{brs1,brs2}, in which the masses of the bosons (except the pion) decrease in a 
nuclear medium due to either normal hadronic many-body effects or the partial 
restoration of chiral symmetry at finite density. In the present work, we examine the 
role of density-dependent corrections to the nuclear interaction due to the 
leading-order chiral three-nucleon force (3NF) at one-loop order. For the 
density-independent two-body part of the interaction we use the low-momentum 
interaction $V_{\rm low-k}$ derived from the Idaho N3LO chiral potential 
\cite{entem1,entem} for cutoffs between $\Lambda_{\rm low-k}$ = 2.1 fm$^{-1}$ and 
2.3 fm$^{-1}$. For densities 
in the neighborhood of saturated nuclear matter ($\rho_0 = 0.16$ fm$^{-3}$, 
we find a large suppression of the 
GT matrix element that is almost entirely due to the short-range component of the 
chiral 3NF. There are three contributions arising from the long-range two-pion 
exchange component of the chiral 3NF, but the largest two terms approximately cancel. 
Moreover, we find that the density-dependent corrections resulting from the medium-range 
3NF are naturally small. We also calculate the Gamow-Teller strengths from the ground 
state of $^{14}$N to the excited states of $^{14}$C and find that they are in 
satisfactory agreement with recent experimental data \cite{negret}, although generally 
they are much less sensitive to the density dependence of the nuclear interaction than 
the ground state to ground state transition.

The present paper is organized as follows. In Section \ref{simni} we develop a model 
for the density-dependent nucleon-nucleon (NN) interaction in-medium. For the 
density-independent part, we use model space/renormalization group techniques to 
construct a low-momentum two-body nuclear interaction based on the N3LO chiral 
NN potential. We then present analytic expressions for the six unique 
density-dependent contributions to the in-medium NN interaction derived from the 
leading-order chiral 3NF at one-loop order. As suggested in ref.\ \cite{nogga}, 
the two low-energy constants, $c_D$ and $c_E$, of the medium- and short-range 
chiral 3NF are chosen to reproduce the binding energies of light nuclei for a 
given low-momentum scale $\Lambda_{\rm low-k}$. In Section 3 we present the 
results of our shell model calculation, and in Section 4 we give a brief summary 
and outlook.

\section{In-medium nuclear interaction}
\label{simni}

\subsection{Free-space low-momentum nucleon-nucleon interaction}
For many years there has been much effort devoted to the construction of 
realistic models of the nuclear interaction, which are important not only for 
understanding the properties of normal nuclei but also for constraining 
the structure of dense stellar objects such as neutron stars. The underlying 
assumption of these models is that the nuclear force arises primarily from 
the exchange of various mesons, though the choice of which mesons to 
include and how to account for the strong short-distance repulsion in the 
NN interaction are largely model-dependent. Nevertheless, 
various models containing between 30 to 40 free parameters (typically meson 
masses, coupling constants, and form factor cutoffs) are able to reproduce all 
of the experimental $pp$ and $pn$ scattering phase shifts below a 
laboratory energy of 350 MeV, as well as the properties of the deuteron, with 
a $\chi^2$ per degree of freedom of about 1.

Recently a program has been developed \cite{bogner02,bogner03} to understand 
the scale-dependence of the NN interaction from the point of view of effective 
field theory and the renormalization group. Since the short-distance behavior 
of the NN interaction is not constrained experimentally, renomalization group 
techniques have been used in refs.\ \cite{bogner02,bogner03} to evolve these 
(bare) NN interactions down to the scale at which our experimental information 
stops, that is, around a center of mass momentum $p \simeq \Lambda_{\rm low-k} 
\simeq 2.1$ fm$^{-1}$. These low-momentum interactions, \vlkn, are phase-shift 
equivalent to the underlying bare interaction up to a predefined cutoff scale 
$\Lambda_{\rm low-k}$. As the decimation scale is reduced to 
$\Lambda_{\rm low-k} \simeq 2.1$ fm$^{-1} \simeq 400$ MeV, all of
these low-momentum interactions flow to a nearly universal interaction. The 
method for constructing such an interaction is explained below.

One begins with the half-on-shell $T$-matrix\footnote{This real-valued 
quantity $T(p',p)$ is often referred to as the $K$-matrix.} for free space 
scattering,
\begin{equation}
T(p',p) = V_{NN}(p',p) + \frac{2}{\pi}{\cal P} \int _0 ^{\infty} 
\frac{V_{NN}(p',q)T(q,p)} {p^2-q^2} q^2 dq,
\end{equation}
from which one defines a low-momentum half-on-shell $T$-matrix by
\begin{equation}
T_{\rm low-k }(p',p) = V_{\rm low-k }(p',p) + \frac{2}{\pi}{\cal P} 
\int _0 ^{\Lambda_{\rm low-k}} \frac{V_{\rm low-k }(p',q) T_{\rm low-k} 
(q,p)}{p^2-q^2} q^2 dq,
\end{equation}
where $\cal P$ denotes the principal value and the cutoff $\Lambda_{\rm low-k}$ 
is a priori arbitrary. To preserve phase shifts, $\tan \delta(p) = -p\, T(p,p)$, 
one requires that these two $T$-matrices be equal for relative momenta less 
than the cutoff $\Lambda_{\rm low-k}$,
\begin{equation}
T_{\rm low-k}(p',p) = T(p',p) \, , \hspace{.1in} {\rm for} \hspace{.2in} p',p 
< \Lambda_{\rm low-k}\, ,
\end{equation}
a condition which defines the low-momentum potential \vlkn. It can be shown 
\cite{bogner02} that a solution to these equations is given by the 
Kuo-Lee-Ratcliff folded diagram effective interaction \cite{klr1,ko90}. There are
several schemes \cite{And,Kren} available for accurately computing 
$V_{\rm low-k}$, and each scheme preserves the deuteron properties. Under
this (scale) decimation procedure, all high-precision NN potentials
flow, as \mbox{$\Lambda_{\rm low-k} \rightarrow 2.1$ $\rm fm^{-1}$}, to a 
nearly unique low-momentum potential $V_{\rm low-k}$. In this study we consistently 
employ the N3LO chiral NN potential in deriving $V_{\rm low-k}$.

Given this model of the nuclear interaction in free space, the question 
remains how to extend the description to a nuclear medium with densities close 
to that of saturated nuclear matter. It is well known that such two-body interactions 
alone are unable to reproduce simultaneously the known saturation energy 
and density of symmetric nuclear matter. The traditional approach is to 
include a three-nucleon force, but for many of the traditional NN interaction 
models it is difficult 
to systematically construct a 3NF that is consistent with the 
underlying two-body interaction. However, by exploiting the separation of scales in the framework of
chiral effective field theory, a systematic and consistent construction of two-, three-, and four-nucleon
forces has become possible (for a recent review, see ref.\ \cite{epelbaum2}). The key element there is a 
power-counting scheme which orders the contributions in powers of small external momenta over the chiral 
symmetry breaking scale. Long-range effects from multi-pion exchanges between nucleons are treated explicitly
(and calculated within chiral perturbation theory), while the short-distance dynamics is encoded in nucleon
contact terms. When applied to two- and few-nucleon problems, these chiral potentials are regulated by exponential
functions \cite{entem, epelbaum2} with cutoffs ranging from 500 to 700 MeV, in order to eliminate (unphysical)
high-momentum components.

The construction of decimated low-momentum three-body forces
consistent with the two-body decimation for \vlk is currently a challenge. A common 
practice is to adjust the parameters of the chiral 3NF so that the binding 
energies of $^3$H, $^3$He, and $^4$He
are reproduced. Such a calculation was carried out in \cite{nogga} using
the Argonne $v_{18}$ \cite{argonne} low-momentum interaction, where in Table \ref{cdcelowk} 
we restate the values of the low-energy constants $c_D$ and $c_E$ determined 
in \cite{nogga} as a 
function of $\Lambda_{\rm low-k}$.
For a cutoff scale of $\Lambda_{\rm low-k} = 2.3$ fm$^{-1}$, \vlk is only
weakly model dependent. Therefore, we expect the values of $c_D$ and
$c_E$ derived for the Argonne $v_{18}$ potential at $\Lambda_{\rm low-k} = 2.3
$ fm$^{-1}$ to be applicable for the low-momentum N3LO chiral potential employed in 
the present paper.
\begin{table}[htb]
\begin{center}
\begin{tabular}{|c||c|c|} \hline
$\Lambda_{\rm low-k}$ & $c_D$ & $c_E$\\ \hline
2.1 fm$^{-1}$ & $-2.062$ & $-0.625$\\ \hline
2.3 fm$^{-1}$ & $-2.785$ & $-0.822$\\ \hline
\end{tabular}
\caption{The values of the low-energy constants $c_D$ and $c_E$ of the chiral 
three-nucleon interaction fit \cite{nogga} to the 
binding energies of $A=3,4$ nuclei for different
values of the momentum cutoff $\Lambda_{\rm low-k}$. The interaction \vlk is
derived from the Argonne $v_{18}$ potential.}
\label{cdcelowk}
\end{center}
\end{table}

\subsection{In-medium nucleon-nucleon interaction}
In this section we derive from the leading-order chiral three-nucleon 
interaction an effective density-dependent in-medium NN interaction. As shown 
in ref.\ \cite{achim} the chiral three-nucleon interaction plays an essential
role in obtaining the saturation of nuclear matter when using the universal 
low-momentum NN potential $V_{\rm low-k}$ in Hartree-Fock calculations. The
parameters of its two-pion exchange component, $c_1 =-0.81\,$GeV$^{-1}$,
$c_3=-3.2\,$GeV$^{-1}$, $c_4 =5.4\,$GeV$^{-1}$, are well determined from
fits to low-energy NN phase shifts and mixing angles \cite{entem}. If one restricts the two large coefficients 
$c_{3,4}$ to their (dominant) $\Delta(1232)$-resonance contributions, the 
celebrated three-nucleon force of Fujita and Miyazawa \cite{fujita} is 
actually recovered. The coupling parameters associated with the
one-pion exchange component $(c_D)$ and the short-range contact-term ($c_E$) of the
chiral 3N-interaction have also been adjusted in refs.\ \cite{nogga,evgeni} to
binding energies of three- and four-nucleon systems ($^3$H, $^3$He, and $^4$He).

We are considering the (on-shell) scattering of two nucleons in the nuclear
medium in the center-of-mass frame, $N_1(\vec p\,)+ N_2(-\vec p\,) \to N_1(\vec 
p+\vec q\,) + N_2(-\vec p-\vec q\,)$, i.e. the total momentum of the
two-nucleon system is zero in the nuclear matter rest frame before and after
the scattering.  The magnitude of the in- and out-going nucleon momenta is $p
= |\vec p\,| = |\vec p+\vec q\,|$, and $q= |\vec q\,|$ gives the magnitude of 
the momentum transfer. Having discussed the  kinematics, we reproduce first 
(for the purpose of comparison) the expression for the (bare) $1\pi$-exchange:
\begin{equation} V_{NN}^{(1\pi)} = - {g_A^2 M_N \over 16 \pi f_\pi^2} \vec \tau_1 \cdot  
\vec \tau_2 \,  {\vec \sigma_1 \cdot \vec q \,\vec \sigma_2 \cdot \vec q\over  
m_\pi^2  + q^2}\,. \end{equation}   
Here, $g_A=1.3$ denotes the nucleon axial-vector coupling constant, $f_\pi = 
92.4\,$MeV is the weak pion decay constant, and $m_\pi = 135\,$MeV is the (neutral) 
pion mass. Furthermore, $\vec \sigma_{1,2}$ and $\vec \tau_{1,2}$ are the
usual spin and isospin operators of the two nucleons. Note that we have included an additional factor
of $M_N/4\pi$ in $V_{NN}$ in order to be consistent with the conventions chosen for \vlkn.

\begin{figure}
\begin{center}
\includegraphics[scale=1.1,clip]{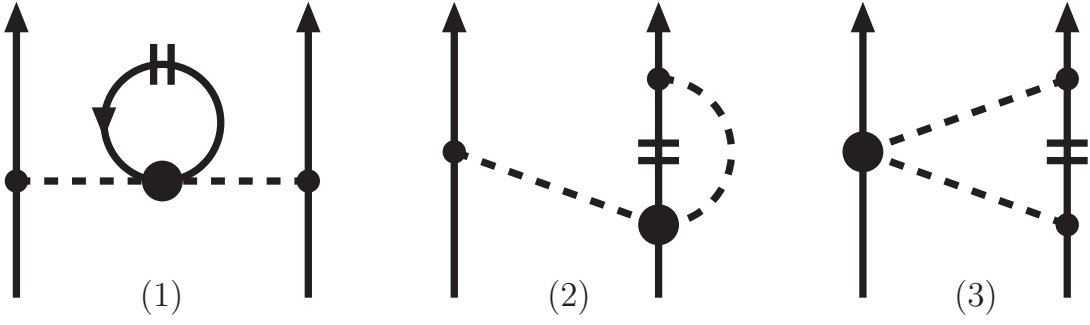}
\end{center}
\vspace{-.5cm}
\caption{In-medium NN interaction generated by the two-pion exchange component
($\sim c_{1,3,4}$) of the chiral three-nucleon interaction. The short 
double-line symbolizes the filled Fermi sea of nucleons, i.e. the medium
insertion $-2\pi \delta(l_0)\, \theta(k_f-|\vec l\,|)$ in the in-medium
nucleon propagator. Reflected diagrams are not shown.}  
\label{mfig1}
\end{figure}

We start with those contributions to the in-medium NN-interaction $V_{NN}^{\rm 
med}$ that are generated by the $2\pi$-exchange component of the chiral
three-nucleon force. The three different topologies for (non-vanishing)
one-loop diagrams are  
shown in Fig.\ \ref{mfig1}. The short double-line on a nucleon propagator symbolizes the 
filled Fermi sea of nucleons, i.e. the medium insertion $-2\pi \delta(l_0)\,
\theta(k_f- |\vec l\,|)$ in the in-medium nucleon propagator. In effect, the
medium insertion  sums up hole propagation and the absence of particle
propagation below the Fermi surface  $|\vec l\,|<k_f$. The left diagram in 
Fig.\ \ref{mfig1} represents a $1\pi$-exchange with a Pauli blocked in-medium 
pion self-energy and the corresponding contribution to $V_{NN}^{(\rm  med)}$ reads:
\begin{equation} V_{NN}^{\rm med,1}= {g_A^2 M_N \rho \over 8 \pi f_\pi^4}\,\vec \tau_1  
\cdot \vec \tau_2 \,{\vec \sigma_1 \cdot\vec q \,\vec \sigma_2 \cdot \vec q 
\over (m_\pi^2 + q^2)^2}\,(2c_1 m_\pi^2 +c_3 q^2)\,.
\label{med1}
\end{equation}
The Fermi momentum $k_f$ is related to the nucleon density in the usual way,
$\rho= 2k_f^3/3\pi^2$. Since $c_{1,3}<0$, this term corresponds to an
enhancement of the bare $1\pi$-exchange. It can be interpreted in terms of the 
reduced (spatial) pion decay constant, $f_{\pi,s}^{*2}=f_\pi^2+2c_3\rho$, in the 
nuclear medium that replaces $f_\pi^2$ in the denominator of eq.(\ref{med1}). The second
diagram in Fig.\ \ref{mfig1} includes vertex corrections 
to the $1\pi$-exchange caused by Pauli blocking in the nuclear medium. The 
corresponding contribution to the in-medium NN-interaction has the form:
\begin{eqnarray} V_{NN}^{\rm med,2}&=& {g_A^2 M_N \over 32\pi^3 f_\pi^4}\vec \tau_1  
\cdot \vec \tau_2 \,  {\vec \sigma_1 \cdot \vec q \,\vec \sigma_2 \cdot \vec q 
\over m_\pi^2 + q^2}\, \bigg\{-4c_1 m_\pi^2 \Big[\Gamma_0(p)+\Gamma_1(p) \Big]
\nonumber \\ && - (c_3+c_4) \Big[q^2 \Big(\Gamma_0(p)+2\Gamma_1(p)+
\Gamma_3(p)\Big)+ 4\Gamma_2(p)\Big] + 4c_4 \bigg[ {2k_f^3 \over
  3}-m_\pi^2\Gamma_0(p)\bigg] \bigg\}\,.
\label{med2}
\end{eqnarray}
Here, we have introduced the functions $\Gamma_j(p)$ which result from Fermi 
sphere integrals over a (static) pion-propagator $[m_\pi^2+(\vec l+\vec p\,
)^2]^{-1}$:
\begin{equation} \Gamma_0(p) = k_f - m_\pi \bigg[ \arctan{k_f+p \over m_\pi} +
\arctan{k_f-p \over m_\pi}\bigg] + {m_\pi^2 +k_f^2 -p^2 \over 4p}\ln {m_\pi^2  
+(k_f+p)^2 \over m_\pi^2+(k_f-p)^2} \,, \end{equation}   
\begin{equation} \Gamma_1(p) = {k_f \over 4p^2}  (m_\pi^2+k_f^2+p^2) -\Gamma_0(p)
-{1\over 16 p^3} \Big[m_\pi^2 +(k_f+p)^2\Big]\Big[ m_\pi^2+(k_f-p)^2\Big]\ln 
{m_\pi^2   +(k_f+p)^2 \over m_\pi^2+(k_f-p)^2} \,, \end{equation} 
 \begin{equation} \Gamma_2(p)= {k_f^3 \over 9}+{1\over6} (k_f^2-m_\pi^2-p^2)
\Gamma_0(p)+{1\over6} (m_\pi^2+k_f^2-p^2)\Gamma_1(p)\,, \end{equation}
\begin{equation} \Gamma_3(p)= {k_f^3 \over 3p^2}-{m_\pi^2+k_f^2+p^2 \over  2p^2}
\Gamma_0(p)-{m_\pi^2+k_f^2+3p^2 \over 2p^2}\Gamma_1(p)\,. \end{equation}
By analyzing the momentum and density dependent factor in eq.(\ref{med2}) relative to
$V_{NN}^{(1\pi)}$, one finds that this contribution corresponds to a reduction
of the $1\pi$-exchange in the nuclear medium. Approximately, this feature can
be  interpreted in terms of a reduced nucleon axial-vector constant
$g_A^*(k_f)$.

The right diagram in Fig.\ \ref{mfig1} represents Pauli blocking effects on chiral 
$2\pi$-exchange. Evaluating it together with the reflected diagram one finds
the following contribution to the in-medium NN-interaction:    
\begin{eqnarray} V_{NN}^{\rm med,3} &=& {g_A^2 M_N \over 64 \pi^3 f_\pi^4}\bigg\{  -12 c_1 
m_\pi^2 \Big[2\Gamma_0(p)- (2m_\pi^2+q^2) G_0(p,q)\Big] \nonumber \\ && -c_3
\Big[8 k_f^3-12(2m_\pi^2+q^2) \Gamma_0(p) -6q^2\Gamma_1(p)+3(2m_\pi^2+q^2)^2 G_0(p,q)
\Big]\nonumber \\&& +4c_4\,\vec \tau_1 \cdot  \vec \tau_2\, (\vec \sigma_1
\cdot \vec \sigma_2\, q^2 - \vec \sigma_1 \cdot  \vec q \,\vec\sigma_2\cdot
\vec q\,) G_2(p,q) \nonumber \\ && -(3c_3+c_4\vec \tau_1 \cdot \vec \tau_2 )\,
i ( \vec \sigma_1 +\vec \sigma_2 )\cdot(\vec q \times \vec p\,)\Big[2\Gamma_0(p)
+2\Gamma_1(p) - (2m_\pi^2+q^2)\nonumber \\ &&\times \Big(G_0(p,q)+2 G_1(p,q)
\Big) \Big] -12 c_1 m_\pi^2\, i ( \vec \sigma_1 +\vec \sigma_2 ) \cdot(\vec q 
\times \vec p\,)  \Big[G_0(p,q)+2 G_1(p,q)\Big] \nonumber \\ && + 4c_4\, 
\vec\tau_1 \cdot \vec \tau_2 \,\vec \sigma_1 \cdot (\vec q \times \vec p\,)\,
\vec\sigma_2 \cdot( \vec q \times \vec p \,) \Big[G_0(p,q) +
4G_1(p,q)+4G_3(p,q) \Big]\bigg\}\,. 
\label{med3}
\end{eqnarray}
One observes that in comparison to the analogous $2\pi$-exchange interaction
in vacuum (see section 4.2 in ref.\ \cite{nnpap}) the Pauli blocking in the 
nuclear medium  has generated additional spin-orbit terms, $i(\vec\sigma_1 
+\vec \sigma_2 
)\cdot(\vec q \times \vec p\,)$, and quadratic spin-orbit terms, $\vec \sigma_1 
\cdot (\vec q \times \vec p\,)\,\vec\sigma_2 \cdot( \vec q \times \vec p \,)$, 
written in the last three lines of eq.(\ref{med3}). The density dependent spin-orbit
terms (scaling with $c_{3,4}$) in the in-medium NN-interaction $V_{NN}^{\rm  med,3}$ 
demonstrate clearly and explicitly the mechanism of three-body induced
spin-orbit forces proposed long ago by Fujita and Miyazawa \cite{fujita}.  

The functions $G_j(p,q)$ appearing in eq.(\ref{med3}) result from Fermi 
sphere
integrals over the product of two  different pion-propagators. Performing the
angular integrations analytically one arrives at:
\begin{equation} G_{0,*,**}(p,q) = {2\over q} \int_0^{k_f}\!\! dl\,  {\{l,l^3,l^5\} 
\over \sqrt{A(p)+q^2l^2} } \ln { q\, l+\sqrt{A(p)+q^2l^2}\over \sqrt{A(p)}}\,,
\end{equation} 
with the abbreviation $A(p)= [m_\pi^2 +(l+p)^2][ m_\pi^2+(l-p)^2]$. The other
functions with $j=1,2,3$ are obtained by solving a system of linear equations:
\begin{eqnarray} G_1(p,q)&=& {\Gamma_0(p)-(m_\pi^2+p^2)G_0(p,q) -G_*(p,q) \over 
4p^2-q^2} \,,\\ G_{1*}(p,q)&=&  {3\Gamma_2(p)+p^2\Gamma_3(p)-(m_\pi^2+p^2)G_*(p,q)
  -G_{**}(p,q) \over  4p^2-q^2} \,,\\G_2(p,q)&=&(m_\pi^2+p^2)G_1(p,q)+G_*(p,q)+
G_{1*}(p,q)\,,\\ G_3(p,q)&=& {{1\over 2}\Gamma_1(p)-2(m_\pi^2+p^2)G_1(p,q) 
-2G_{1*}(p,q) -G_{*}(p,q) \over 4p^2-q^2} \,.\end{eqnarray} 
In this chain of equations the functions indexed with an asterisk play 
only an auxiliary role for the construction of $G_{1,2,3}(p,q)$. We note that all 
functions $G_j(p,q)$ are non-singular at $q=2p$ (corresponding to scattering
in backward direction).  
\begin{figure}
\begin{center}
\includegraphics[scale=1.1,clip]{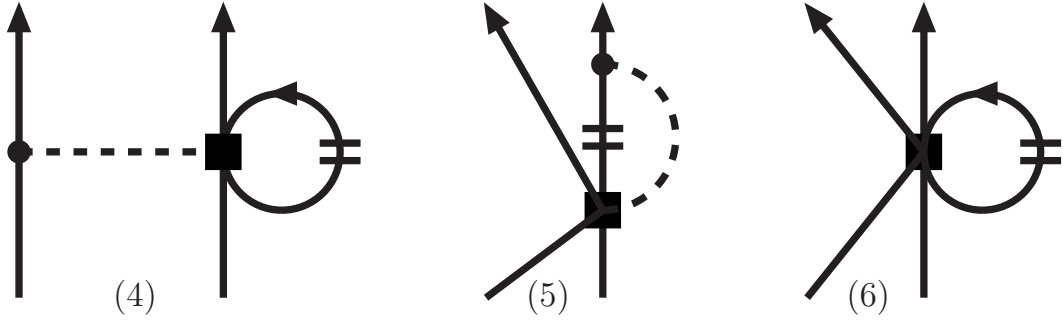}
\end{center}
\vspace{-.5cm}
\caption{In-medium NN interaction generated by the one-pion exchange ($\sim 
c_D$) and short-range component ($\sim c_E$) of the chiral three-nucleon
interaction.} 
\label{mfig2}
\end{figure}

Next, we come to the $1\pi$-exchange component of the chiral three-nucleon
interaction proportional to the parameter $c_D/\Lambda_{\chi}$, where $c_D \simeq -2$
for a scale of $\Lambda_{\chi} =0.7\,$GeV \cite{nogga}. The filled black square in
the first and second diagram of Fig.\ \ref{mfig2} symbolizes the corresponding 
two-nucleon one-pion contact interaction. By closing a nucleon line at the
contact vertex, one obtains a vertex correction (linear in the density $\rho$) 
to the  $1\pi$-exchange:
\begin{equation} V_{NN}^{\rm med,4} = -{g_A M_N c_D \rho \over 32 \pi f_\pi^4\Lambda_{\chi}}
\,\vec \tau_1 \cdot \vec \tau_2 \,{\vec \sigma_1 \cdot\vec q \,\vec \sigma_2 
\cdot \vec q  \over  m_\pi^2 + q^2} \,.\end{equation}
Since $c_D$ is negative, this term reduces again the bare $1\pi$-exchange,
roughly by about $16\%$ at normal nuclear matter density $\rho_0=0.16\,
$fm$^{-3}$. The second diagram in Fig.\ \ref{mfig2} includes Pauli blocked (pionic) 
vertex corrections to the short-range NN interaction. The corresponding 
contribution to the density dependent in-medium NN interaction reads:
\begin{eqnarray} V_{NN}^{\rm med,5}&=& {g_A M_N c_D\over 64 \pi^3 f_\pi^4\Lambda_{\chi}}\vec 
\tau_1 \cdot \vec \tau_2 \bigg\{2 \vec \sigma_1 \cdot \vec \sigma_2\,\Gamma_2(p)
 +\bigg[\vec \sigma_1 \cdot \vec \sigma_2 \bigg( 2p^2-{q^2\over 2}\bigg) + \vec 
\sigma_1 \cdot \vec q\,\vec\sigma_2  \cdot \vec q\, 
\bigg(1-{2p^2\over q^2}\bigg) \nonumber \\ &&  -{2\over q^2}\,\vec\sigma_1 \cdot (\vec q \times 
\vec p\,)\,\vec\sigma_2 \cdot(\vec q\times \vec p \,)\bigg]
\Big[\Gamma_0(p)+2\Gamma_1(p)+\Gamma_3(p)\Big] \bigg\}\,, 
\label{med5}
\end{eqnarray}
where we have used an identity for $\vec \sigma_1 \cdot \vec p\,\, \vec\sigma_2 
\cdot\vec p+ \vec \sigma_1 \cdot (\vec p+\vec q\,)\, \vec\sigma_2\cdot (\vec p+
\vec q\,)=[\dots]$. The ellipses stands for the combination of spin operators 
written in the square bracket of eq.(\ref{med5}).  

Finally, there is the short-range component of the chiral 3N interaction, 
represented by a three-nucleon contact-vertex proportional to  $c_E/\Lambda_{\chi}$. By
closing one nucleon line (see right diagram in Fig.\ 2) one obtains the
following contribution to the in-medium NN-interaction:
\begin{equation}  V_{NN}^{\rm med,6} =-{3 M_N c_E \rho \over 8 \pi f_\pi^4\Lambda_\chi} \,,
\end{equation}
which simply grows linearly in density $\rho=2k_f^3/3\pi^2$ and is independent of 
spin, isospin and nucleon momenta.\footnote{In order to facilitate the 
computation of symmetry factors and spin and isospin traces, we have modeled 
(for that purpose)
the three-nucleon contact-interaction by heavy isoscalar boson exchanges.}

For implementation of the in-medium NN-interaction into nuclear structure 
calculations, the matrix elements of $V_{NN}^{\rm med}$ in the $LSJ$ basis are 
needed. These are obtained by setting $\vec \tau_1 \cdot \vec \tau_2 = 4T-3$, 
with $T=0,1$ the total isospin, and $q = p \sqrt{2(1-z)}$ and performing a 
projection $\int_{-1}^{1} dz\, P_L(z)\dots$ with Legendre polynomials (for 
details see section 3 in ref.\ \cite{nnpap}). The resulting diagonal 
spin-singlet ($S=0$, $L=J$) and diagonal spin-triplet ($S=1$, $L=J-1,J,J+1$) 
matrix elements as well as the off-diagonal mixing matrix elements ($S=1$, 
$L=J-1$, $L'=J+1$) are then functions of the momentum $p$ and $k_f$ (or 
equivalently the nucleon density $\rho$). Note that the off-diagonal mixing 
matrix elements arise exclusively from the tensor
operator  $\vec \sigma_1 \cdot  \vec q \,\vec\sigma_2\cdot \vec q$ and the 
quadratic spin-orbit operator  $\vec \sigma_1 \cdot (\vec q \times \vec p\,)\,
\vec\sigma_2 \cdot( \vec q \times \vec p \,)$.

\begin{figure}
\begin{center}
\includegraphics[height=15cm,angle=270]{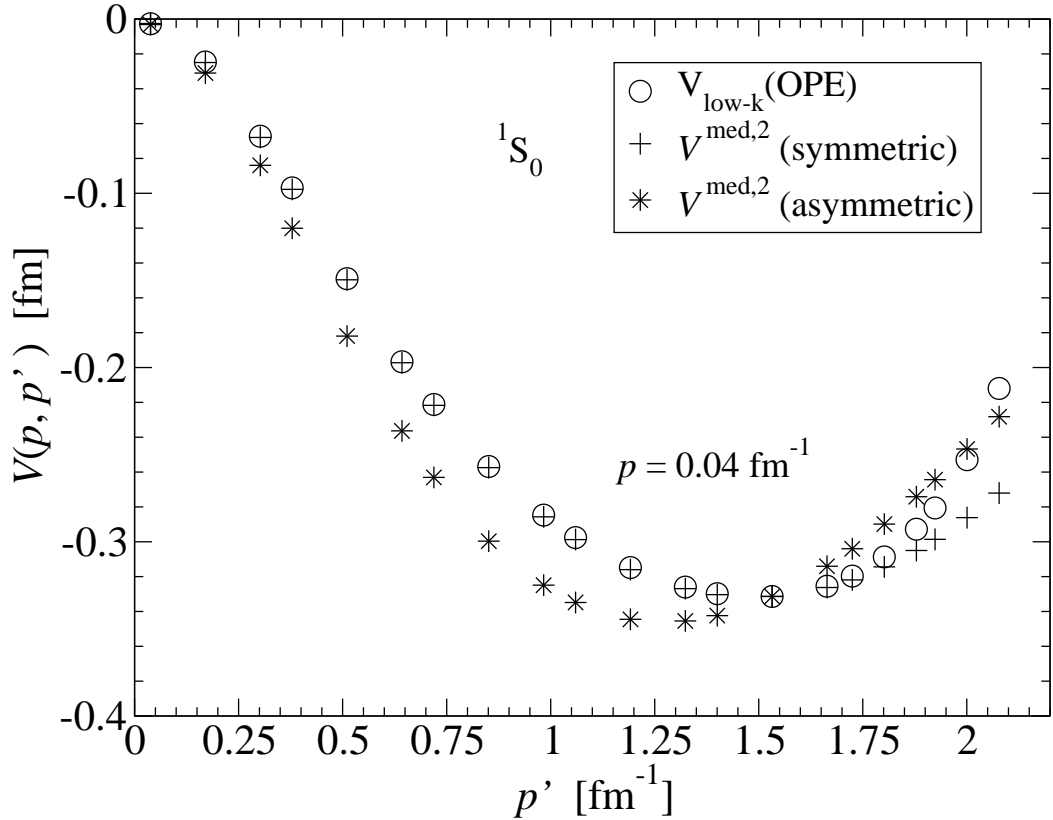}
\end{center}
\vspace{-.5cm}
\caption{Two different off-shell extrapolations for the density dependent $V^{\rm med,2}_{NN}$ at $\rho =\rho_0$ compared to the off-shell dependence of the low-momentum one-pion-exchange in the $^1S_0$ partial-wave. Matrix elements are shifted and scaled for comparison purposes (see text for details).}
\label{offshell}
\end{figure}

The restriction to on-shell amplitudes greatly simplifies the calculation of
the medium-dependent corrections to the nuclear interaction. However, in
calculating the shell model matrix elements one must know the components
of the interactions also off-shell. That is, one needs to consider $N_1(\vec p\,)+ N_2(-\vec p\,) \to N_1(\vec p+\vec q\,) + N_2(-\vec p-\vec q\,)$, where $p=|\vec{p}\, |$ and $p^\prime = |\vec{p}+\vec{q}\, | \neq p$. Rather than calculate the full off-shell in-medium NN interaction, which would give rise to new spin-dependent operators beyond those present in free-space two-nucleon scattering, we look at two prescriptions for extending the on-shell amplitudes off-shell. First, we consider the symmetric extrapolation obtained by replacing $p^2$ in the above on-shell expressions with $\frac{1}{2}(p^2+{p^\prime}^2)$. Second, we consider the asymmetric extrapolation obtained by simply introducing no explicit dependence on ${p^\prime}^2$. As we discuss later, the largest off-shell dependence comes from $V^{\rm med,2}_{NN}$. In Fig.\ \ref{offshell} we compare the two off-shell extrapolations for $V^{\rm med,2}_{NN}$ at $\rho_0 = 0.16$ fm$^{-3}$ in the $^1S_0$ partial wave to the off-shell dependence of the one-pion-exchange (OPE) low-momentum interaction, where $\Lambda_{\rm low-k} = 2.1$ fm$^{-1}$. We have fixed the value of $p=0.04$ fm$^{-1}$, and for ease of comparison among the three sets of off-diagonal matrix elements, we have changed the overall sign of the OPE interaction, shifted all three to the origin, and scaled the two medium-dependent interactions so that they coincide with the low-momentum OPE at $p^\prime = 1.6$ fm$^{-1}$. We find that there is excellent agreement between the symmetric off-shell extrapolation and the low-momentum OPE for momenta up to nearly $p' = 1.75$ fm$^{-1}$. Therefore, throughout the remainder of this study we employ only the symmetric off-shell extrapolation. This completes our construction of the density-dependent in-medium NN interaction arising from the leading-order chiral three-nucleon force.

\section{Shell model calculation}
\subsection{Formalism}
Nowadays there are several computationally intense {\it ab initio} approaches to
solve the nuclear many-body problem for finite nuclei. These include Green
function Monte Carlo techniques \cite{pieper}, the no-core shell model 
\cite{navratil}, and coupled cluster methods \cite{heisen1, dean}. However,
$A=14$ nuclei are currently still too complicated for any of these methods to
treat exactly. Therefore, in this study we employ the standard shell model,
which restricts the allowable configurations to $0\hbar \omega$ but includes
the mixing of more complicated configurations through perturbation theory. 
Indeed, the shell model is expected to provide a good description of light nuclei close to a doubly-magic nucleus (e.g., $^{16}$O in the present study).

We describe the states of $^{14}$C and $^{14}$N as consisting of two $0p$
holes in a closed $^{16}$O core. The harmonic oscillator parameter is chosen
to be $\hbar \omega = 14$ MeV, which yields good agreement with the experimental charge
distribution of $^{16}$O. The second parameter in the model is the energy
splitting between the two $0p$ orbitals, $\epsilon = e(p_{1/2})- e(p_{3/2})=6.4$ MeV, which we have taken from the experimental excitation energy of the first $3/2^-$ state of $^{15}$N located 6.4 MeV above the $1/2^-$ ground state.

\begin{figure}[hbt]
\centering
\includegraphics[height=6cm]{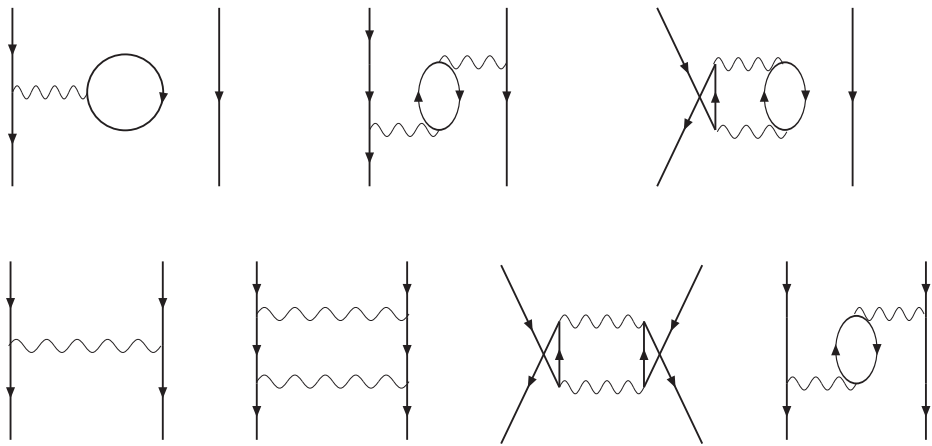}
\caption{Diagrams contributing to the effective interaction $V_{\rm eff}$ in our present calculation. Wavy lines represent the density-dependent in-medium nuclear interaction caculated in Section \ref{simni}.}
\label{qbox}
\end{figure}

To obtain the ground state energies and wavefunctions we construct the shell model effective interaction following the formalism explained in ref.\ \cite{ko90}. 
The full nuclear many-body problem
\be
H\Psi _n=E_n \Psi _n,
\ee
which presently cannot be solved for mass number $A=14$, is replaced with a model space problem
\be
H_{\rm eff}\chi _m=E_m \chi _m.
\ee
In this equation
\be
H=H_0+V \hspace{.1in} {\rm and} \hspace{.1in} H_{\rm eff}=H_0+V_{\rm eff},
\ee
where $H_0$ is equal to the sum of the kinetic energy and single-particle harmonic oscillator potential, $E_n=E_n(A=14)-E_0(A=16,\rm core)$, and $V$ denotes
the input NN interaction (in this case, it is provided by the density-dependent potential $V_{\rm low-k} + V^{\rm med}_{NN}$). The effective interaction $V_{\rm eff}$ is obtained from the folded-diagram formalism detailed in ref.\ \cite{jensen95}. In the
present study, the $\hat{Q}$-box \cite{jensen95} is calculated using hole-hole irreducible diagrams of first- and
second-order in $V_{\rm low-k} + V^{\rm med}_{NN}$ as shown in Fig.\ \ref{qbox}. To distinguish these many-body particle-hole diagrams from the pion-exchange diagrams contributing to the in-medium NN interaction calculated in Section \ref{simni}, we symbolize the input interaction with a wavy line. In previous work \cite{jensen95}, the strong short-distance repulsion in the NN $S$-wave interaction was mitigated by constructing the $G$-matrix, thereby yielding an effective interaction suitable for perturbation theory techniques. Our present use of low-momentum interactions achieves the same purpose and obviates the construction of the $G$-matrix. Previous studies \cite{achim,jason,cora07} have found that in general low-momentum interactions are suitable for perturbative calculations. In all of these references satisfactory converged results were obtained including terms only up to second order in \vlkn. Moreover, low-momentum interactions are energy independent and do not suffer from possible double counting problems that can arise with $G$-matrix interactions.

We denote the single-hole states by
\begin{equation}
\left | 1 \right > = \left | 0s_{1/2}^{-1} \right >, 
\left | 2 \right > = \left | 0p_{3/2}^{-1} \right >,
\left | 3 \right > = \left | 0p_{1/2}^{-1} \right >,
\end{equation}
and the two-hole states coupled to good angular momentum $J$, isospin $T$, and parity $\pi$ are denoted by $\left | \alpha \beta; J^\pi T \right >$, where $\alpha,\beta = $ 2 or 3. The matrices to diagonalize are
\begin{equation}
\left[ \begin{array}{ccc}
 & \vdots & \\
\cdots & \langle \alpha \beta; 1^+ 0|V_{\rm eff}|\gamma \delta; 1^+ 0 \rangle & \cdots\\
 & \vdots & \end{array} \right] 
+ \left[\begin{array}{ccc}
0 & 0 & 0 \\
0 & -\epsilon & 0 \\
0 & 0 & -2\epsilon \end{array} \right],
\label{matrices}
\end{equation}
for the $J^\pi = 1^+$, $T=0$ states, and a similar $2\times 2$ matrix for 
the $J^\pi = 0^+$, $T=1$ states. The ground states for $^{14}$C and $^{14}$N
are respectively written as\footnote{We note that the $jj$-coupling scheme employed in the present paper differs from the previous density-dependent study \cite{holt} in which the two-particle states were $LS$-coupled.}
\begin{eqnarray}
\psi_i &=& a \left |22;0^+ 1 \right > + b\left|33;0^+ 1\right > \, ,\nonumber \\
\psi_f &=& x \left |22;1^+ 0\right > + y\left|23;1^+ 0\right > + 
z\left|33;1^+ 0\right> \, ,
\label{ls}
\end{eqnarray}
and the reduced Gamow-Teller matrix element $M_{\rm GT}$ is evaluated to be
\begin{equation}
M_{GT} = \sum_{k} \left<\psi_f||\sigma(k)\tau_+(k)||\psi_i\right> = \frac{1}{\sqrt{6}}\left(-2\sqrt{5}ax + 2\sqrt{2}ay + 4by + bz \right ).
\label{gte}
\end{equation}
The half-life $T_{1/2}$ is inversely proportional to the square of the Gamow-Teller 
matrix element, and for the $^{14}$C decay one has
\begin{equation}
T_{1/2}=\frac{1}{f(Z,E_0)}\frac{2\pi^3 \hbar^7 {\rm ln} 2}{m_e^5 c^4 G_V^2}\frac{1}{{g_A^*}^2|M_{GT}|^2}\, ,
\end{equation}
where $E_0=156$ keV is the maximum kinetic energy of the emitted 
electron, $f(Z,E_0)$ is the Fermi integral over phase space, $G_V =G_F \cos \theta_c$ 
is the weak vector coupling constant, and $g_A^*$ is the in-medium axial vector 
coupling constant. Although 
the small end-point energy $E_0$ of the $^{14}$C beta decay 
suppresses the transition by a factor of several hundred, this is not nearly 
sufficient to account for the observed lifetime. In order to obtain the 
anomalously long half-life of $T_{1/2} \simeq 5730$ years, the GT matrix element must be 
anomalously small, $M_{GT} \simeq 2 \times 10^{-3}$.

\subsection{Results}
As a preliminary step, we first calculate the $^{14}$C and $^{14}$N wavefunctions, as well as the Gamow-Teller
transition matrix element, using the bare \vlk derived from the Idaho N3LO potential at a cutoff scale $\Lambda_{\rm low-k}= 2.1$ fm$^{-1}$. We find
\begin{eqnarray}
\left |J^\pi = 0^+, T=1 \right >_1 &=& 0.395 \left |22;0^+,1\right > + 
0.919 \left |33;0^+,1\right > \, ,\nonumber \\
\left |J^\pi = 0^+, T=1 \right >_2 &=& 0.919 \left |22;0^+,1\right > - 
0.395 \left |33;0^+,1\right > \, ,
\label{c14wfn}
\end{eqnarray}
for the two $0^+$ states of $^{14}$C with energy splitting $\Delta E (0^+) = 12.9$ MeV. For the two lowest $1^+$ states of $^{14}$N we find
\begin{eqnarray}
\left |J^\pi = 1^+, T=0 \right >_1 &=& 0.137 \left |22;0^+,1\right > - 
0.676 \left |33;0^+,1\right > + 0.725 \left |33;0^+,1\right > \, ,\nonumber \\
\left |J^\pi = 1^+, T=0 \right >_2 &=& 0.360 \left |22;0^+,1\right > - 
0.670 \left |33;0^+,1\right > - 0.649 \left |33;0^+,1\right >,
\label{n14wfn}
\end{eqnarray}
where the first excited state lies $\Delta E (1^+) = 2.77$ MeV above the ground state. The second excited $J^\pi =1^+, T=0$ state lies nearly 16 MeV higher in energy than the ground state and will therefore be neglected throughout. With these wavefunctions, the ground state to ground state Gamow-Teller transition matrix 
element is found to be
\begin{equation}
M_{GT} = -0.88\, ,
\label{gtv}
\end{equation}
which is much too large to describe the known lifetime of $^{14}$C.

\setlength{\tabcolsep}{.075in}
\begin{table}[htb]
\begin{center}
\begin{tabular}{|c|c|c|c|} 
\multicolumn{4}{c}{$J^\pi = 0^+, T=1$ \hspace{.2in} 
($\rho=\rho_0/10$)} \\ \hline
& $\langle 22| V^{\rm med}_{NN}|22\rangle$ & $\langle22|V^{\rm med}_{NN} |33\rangle$ & $\langle33|V^{\rm med}_{NN} |33\rangle$  \\ \hline
1 &  0.277 &  0.053 &  0.240  \\ \hline
2 & -0.292 & -0.071 & -0.242  \\ \hline
3 &  0.043 & -0.048 &  0.077  \\ \hline
4 & -0.054 & -0.011 & -0.047  \\ \hline
5 &  0.041 &  0.027 &  0.022  \\ \hline
6 &  0.171 &  0.121 &  0.085  \\ \hline
\multicolumn{4}{l}{} \\
\multicolumn{4}{c}{$J^\pi = 0^+, T=1$ \hspace{.2in} 
($\rho=\rho_0$)} \\ \hline
& $\langle 22| V^{\rm med}_{NN}|22\rangle$ & $\langle 22|V^{\rm med}_{NN} |33\rangle$ & $\langle 33|V^{\rm med}_{NN} |33\rangle$  \\ \hline
1 &  2.766 &  0.525 &  2.395  \\ \hline
2 & -3.367 & -0.714 & -2.863  \\ \hline
3 &  1.231 &  0.004 &  1.228  \\ \hline
4 & -0.545 & -0.107 & -0.469  \\ \hline
5 &  0.458 &  0.316 &  0.235  \\ \hline
6 &  1.707 &  1.207 &  0.853  \\ \hline
\end{tabular}
\caption{Matrix elements (in units of MeV) between $0p^{-2}$ states coupled to $(J^\pi,T) = (0^+,1)$ for the six density-dependent contributions to the in-medium NN
interaction at the two densities $\rho_0/10$ and $\rho_0$.}
\label{tnf0}
\end{center}
\end{table}

We now discuss the role of the six density-dependent components $V^{\rm med}_{NN}$ derived in
Section 2 that arise from the lowest-order chiral 3NF. Indeed, since $A=14$ nuclei lie just below a double shell closure, we expect the average density experienced by a valence $p$-shell nucleon to be close to that of saturated nuclear matter. To begin we consider
only the matrix elements of these interactions within the $0p^{-2}$ model 
space. These are shown in Tables \ref{tnf0} and \ref{tnf1} for each of the
six $V^{{\rm med}, i}_{NN}$ at the
two densities $\rho = \rho_0/10$ and $\rho = \rho_0$.
The matrix elements connecting states outside the model space contribute at
second order in the diagrammatic expansion of the effective shell-model
interaction, $V_{\rm eff}$, and will be included later.
\setlength{\tabcolsep}{.075in}
\begin{table}[htb]
\begin{center}
\begin{tabular}{|c|c|c|c|c|c|c|} 
\multicolumn{7}{c}{$J^\pi = 1^+, T=0$ \hspace{.2in}
($\rho=\rho_0/10$)} \\ \hline
& $\langle 22| V^{\rm med}_{NN}|22 \rangle$ & $\langle 23|V^{\rm med}_{NN} |22\rangle$ & $\langle 23|V^{\rm med}_{NN} |23\rangle$ & $\langle 23|V^{\rm med}_{NN} |33\rangle$ & $\langle 22|V^{\rm med}_{NN} |33\rangle$ & 
$\langle 33|V^{\rm med}_{NN} |33\rangle$  \\ \hline
1 &  0.299 & -0.051 &  0.128 &  0.143 & -0.121 &  0.240 \\ \hline
2 & -0.257 &  0.091 & -0.147 & -0.101 &  0.151 & -0.242 \\ \hline
3 &  0.056 &  0.028 & -0.019 & -0.015 & -0.100 &  0.182 \\ \hline
4 & -0.054 &  0.012 & -0.027 & -0.023 &  0.025 & -0.047 \\ \hline
5 &  0.034 & -0.026 &  0.012 &  0.021 & -0.040 &  0.034 \\ \hline
6 &  0.103 & -0.107 &  0.171 & -0.002 & -0.055 &  0.089 \\ \hline
\multicolumn{7}{l}{} \\
\multicolumn{7}{c}{$J^\pi = 1^+, T=0$ \hspace{.2in}
($\rho=\rho_0$)} \\ \hline
& $\langle 22| V^{\rm med}_{NN} |22\rangle$ & $\langle 23|V^{\rm med}_{NN} |22\rangle $ & $\langle 23|V^{\rm med}_{NN} |23\rangle$ & $\langle 23|V^{\rm med}_{NN} |33\rangle$ & $\langle 22|V^{\rm med}_{NN} |33\rangle$ & 
$\langle 33|V^{\rm med}_{NN} |33\rangle$  \\ \hline
1 &  2.994 & -0.505 &  1.282 &  1.427 & -1.209 &  2.395 \\ \hline
2 & -3.193 &  0.865 & -1.718 & -1.291 &  1.589 & -2.863 \\ \hline
3 &  0.634 & -0.464 &  0.524 & -0.035 & -0.927 &  1.255 \\ \hline
4 & -0.542 &  0.123 & -0.273 & -0.230 &  0.247 & -0.469 \\ \hline
5 &  0.322 & -0.286 &  0.319 &  0.100 & -0.276 &  0.303 \\ \hline
6 &  1.027 & -1.074 &  1.714 & -0.016 & -0.550 &  0.885 \\ \hline
\end{tabular}
\caption{Matrix elements (in units of MeV) between $0p^{-2}$ states coupled to $(J^\pi,T) = (1^+,0)$ for the six density-dependent contributions to the NN
interaction in-medium at the two densities $\rho_0/10$ and $\rho_0$.}
\label{tnf1}
\end{center}
\end{table}
In general, the largest matrix elements are those arising from 
$V_{NN}^{\rm med,1}$ and $V_{NN}^{\rm med,2}$, which are respectively the in-medium pion self-energy and vertex correction resulting from the long-range chiral
3NF. However, we find that to a good approximation they cancel. The same is 
true for $V_{NN}^{\rm med,4}$ and $V_{NN}^{\rm med,5}$, which are relatively small
to begin with. Therefore, we would expect $V_{NN}^{\rm med,3}$ and $V_{NN}^{\rm med,6}$
to be the most important density-dependent corrections to the in-medium NN 
interaction for the observables considered here.

\begin{figure}[hbt]
\begin{center}
\includegraphics[height=16cm,angle=-90]{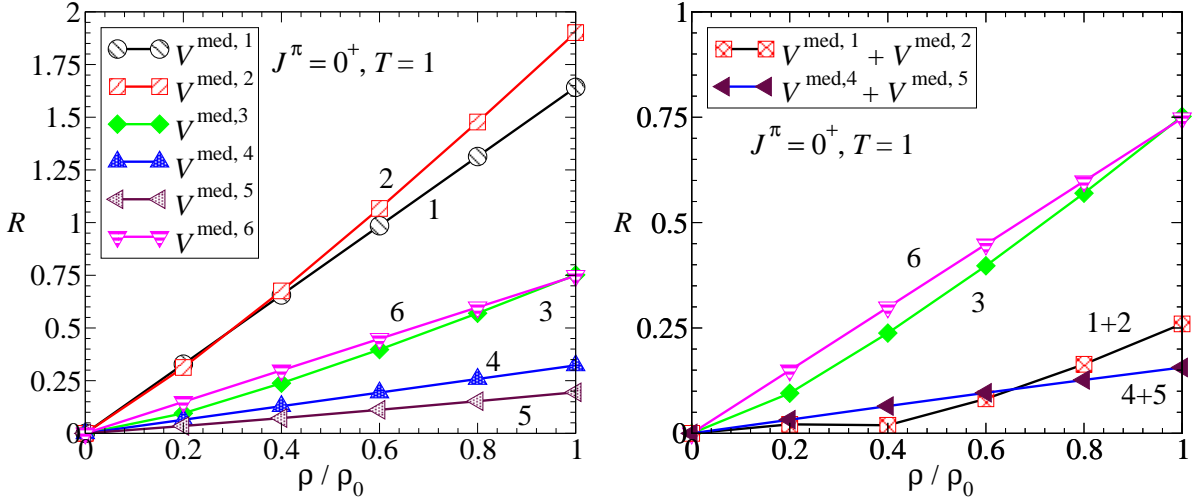}
\caption{The average strength of the three $\left < \alpha \beta; 0^+ 1 \right | V^{\rm med, i}_{NN} \left | \gamma \delta; 0^+ 1\right >$ matrix elements for the six density-dependent contributions to the in-medium NN interaction relative to $V_{\rm low-k}$. The low-momentum interaction is constructed from the Idaho N3LO potential for $\Lambda_{\rm low-k}=2.1$ fm$^{-1}$.}
\label{c14mer0p}
\end{center}
\end{figure}

\begin{figure}[hbt]
\begin{center}
\includegraphics[height=16cm,angle=-90]{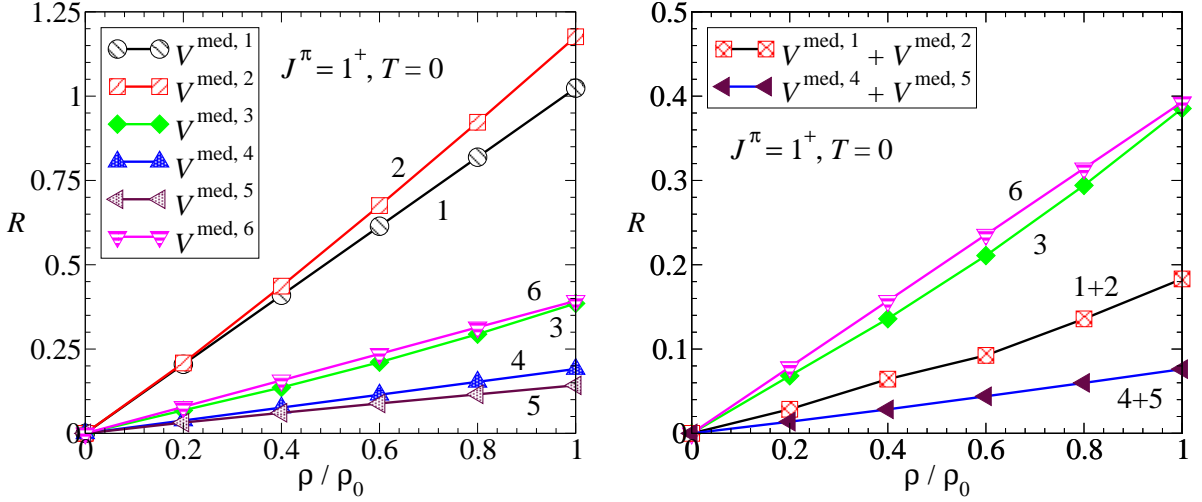}
\caption{The average strength of the six $\left < \alpha \beta; 1^+ 0 \right | V^{\rm med, i}_{NN} \left | \gamma \delta; 1^+ 0\right >$ matrix elements for the six density-dependent contributions to the in-medium NN interaction relative to $V_{\rm low-k}$. The low-momentum interaction is constructed from the Idaho N3LO potential for $\Lambda_{\rm low-k}=2.1$ fm$^{-1}$.}
\label{c14mer1p}
\end{center}
\end{figure}

These observations are made more concrete in the following. As a measure of
the strength of the various components contributing to the
density-dependent part of the NN interaction, we define the following quantity
\begin{equation}
R^i(J^\pi, T, \rho) = \sum_{\alpha \beta \gamma \delta}\frac{1}{n}\left | \frac{\left < \alpha \beta J^\pi T | 
V^{{\rm med},i}_{NN} | \gamma \delta J^\pi T \right > }{\left < \alpha \beta J^\pi T | V_{\rm low-k} | \gamma \delta J^\pi T \right > }
\right |,
\end{equation}
where $\rho$ is the nuclear density. The sum over $\alpha, \beta, \gamma,$ and $\delta$ runs over all possible
configurations with the allowed spin, isospin, and parity, and $n$ is the
number of such configurations (e.g., $n=3$ for the $J^\pi = 0^+, T=1$ states). We
derive \vlk from the chiral N3LO NN interaction and choose 
$\Lambda_{\rm low-k} = 2.1$ fm$^{-1}$. A value of $R^i$ close to 1 indicates that the relevant matrix elements for a particular medium correction are on average equal in magnitude to the matrix elements of $V_{\rm low-k}$ in the given spin-isospin state. We show on the left side of Figs.\ \ref{c14mer0p} and \ref{c14mer1p} the value of $R$ for each of the six density-dependent components of the in-medium nuclear interaction for densities between $\rho = 0$ and $\rho=\rho_0$. The plots on the right hand side of the two figures include also the ratio $R$ for $V^{\rm med,1} + V^{\rm med,2}$ and $V^{\rm med,4} + V^{\rm med,5}$. We see that $V^{\rm med,1}_{NN} + V^{\rm med,2}_{NN}$ as well as $V^{\rm med,4}_{NN} + V^{\rm med,5}_{NN}$ largely cancel, so that the most important contributions from the chiral 3NF are the Pauli-blocked $2\pi$ exchange and the three-nucleon contact interaction proportional to $c_E$.

Before discussing the results of the full calculation shown at the end of this section, we first obtain a physical understanding of the role played by the two dominant components, $V^{\rm med,3}$ and $V^{\rm med,6}$, from lowest-order perturbation theory. Here the density-dependent components induce small changes in the ground state wavefunctions by mixing in the first excited states. This is justified on the basis that the relevant matrix elements of the density-dependent interactions are small compared to the lowest-order matrix elements from $V_{\rm low-k}$ as seen in Figs.\ \ref{c14mer0p} and \ref{c14mer1p}. We include explicitly the higher-order many-body diagrams shown in Fig.\ \ref{qbox} at the end of this section.

Given the ground state wavefunctions obtained from $V_{\rm low-k}$ alone, as shown in eqs.\ (\ref{c14wfn}) and (\ref{n14wfn}), we see that in eq.(\ref{gte}) the first three terms contributing to
the GT matrix element enter with the same sign. It is only the last contribution ($bz$) that reduces the strength of the transition matrix element. In order to suppress the GT transition, the density-dependent corrections must therefore shift strength to the $\left |33\right >$ components of the wavefunctions. A straightforward calculation using the matrix elements shown in Tables \ref{tnf0} and \ref{tnf1} for $\rho = \rho_0$ gives for the un-normalized perturbed wavefunctions arising from $V^{\rm med, 3}_{NN}$
\begin{eqnarray}
\psi^{(1)}_0(0^+,1) &=& \psi^{(0)}_0(0^+,1) - 0.0003 \psi^{(0)}_1(0^+,1)\, , \nonumber \\
\psi^{(1)}_0(1^+,0) &=& \psi^{(0)}_0(1^+,0) + 0.002 \psi^{(0)}_1(1^+,0)\, ,
\label{pert3}
\end{eqnarray}
and similarly from $V^{\rm med, 6}_{NN}$
\begin{eqnarray}
\psi^{(1)}_0(0^+,1) &=& \psi^{(0)}_0(0^+,1) - 0.088 \psi^{(0)}_1(0^+,1)\, , \nonumber \\
\psi^{(1)}_0(1^+,0) &=& \psi^{(0)}_0(1^+,0) - 0.313 \psi^{(0)}_1(1^+,0)\, .
\label{pert6}
\end{eqnarray}
This leading perturbative calculation shows the following features. First, the 
effects due to $V^{\rm med, 6}_{NN}$ dominate those from $V^{\rm med, 3}_{NN}$, and second, one expects a reduction in the GT transition strength, since the effect of $V_{NN}^{{\rm med},6}$ is to increase the strength of the $\left | 33 \right >$ components of both ground state wavefunctions.

We now consider the results of the full calculation, including all density-dependent contributions to second-order in the shell model effective 
interaction shown in Fig.\ \ref{qbox}. In Table \ref{jjtab2} we show the results of our calculations for the 
expansion coefficients of the ground state wavefunctions of $^{14}$C and $^{14}
$N, as well as the reduced GT matrix element, as a function of the nuclear 
density up to $\rho = 1.25\rho_0$.
\setlength{\tabcolsep}{.075in}
\begin{table}[htb]
\begin{center}
\begin{tabular}{|c|c|c|c|c|c|c|} \hline
$\rho/\rho_0$ & $a$ & $b$ & $x$ & $y$ & $z$ & $M_{\rm GT}$ \\ \hline
0.00 & 0.40 & 0.92 & 0.14 & -0.68 & 0.72 & -0.877 \\ \hline
0.25 & 0.37 & 0.93 & 0.13 & -0.67 & 0.73 & -0.833 \\ \hline
0.50 & 0.34 & 0.94 & 0.11 & -0.63 & 0.77 & -0.684 \\ \hline
0.75 & 0.30 & 0.95 & 0.09 & -0.57 & 0.82 & -0.488 \\ \hline
1.00 & 0.25 & 0.97 & 0.07 & -0.49 & 0.87 & -0.267 \\ \hline
1.25 & 0.19 & 0.98 & 0.05 & -0.41 & 0.91 & -0.045 \\ \hline
\end{tabular}
\caption{The coefficients of the $jj$-coupled wavefunctions defined in eq.\
  (\ref{ls}) and the associated reduced GT matrix element as a function of the nuclear
  density $\rho$.}
\label{jjtab2}
\end{center}
\end{table}
We see that the main effect of including the density-dependent modifications of the NN interaction is to strongly suppress the GT transition, as we expected from our previous perturbative analysis. However, in this calculation with theoretical errors neglected, it appears that only at densities above that of saturated nuclear matter would the suppression be strong enough to reproduce the experimentally observed half-life.

The Gamow-Teller transition strength, $B(GT)$, is related to the reduced GT transition matrix element by 
\begin{equation}
B(GT)=(g_A^*)^2 \frac{1}{2J_i+1}|M_{\rm GT}|^2 \simeq |M_{\rm GT}|^2,
\end{equation}
where $J_i =0$ and we have used the approximation that the in-medium 
axial vector coupling constant $g_A^* \simeq 1$. In general, the effective Gamow-Teller operator 
has additional terms in a nuclear medium \cite{arima,castel}:
\begin{equation}
{\cal \vec{O}}_{\rm GT,eff} = g_{LA} \vec{L} + g_A^* \, \vec{\sigma} + g_{PA} \,[Y_2,\vec{\sigma}],
\end{equation}
where $\vec{L}$ is the orbital angular momentum operator and $Y_2$ denotes the rank-2 spherical harmonic $Y_{2,m}$. However, it is well known from theoretical calculations \cite{towner} of the effective GT operator in-medium, as well as beta decay calculations \cite{babrown} performed with phenomenological shell model effective interations, that $g_{LA}$ and $g_{PA}$ are almost negligible and that in light nuclei $g_A^*$ is smaller by 15-20\% 
from its free space value of $g_A = 1.27$ (measured in neutron beta decay). In general, one must calculate both the effective interaction $V_{\rm eff}$ and the effective Gamow-Teller operator ${\cal \vec{O}}_{\rm GT, eff}$. In our calculations we assume a 20\%
reduction of $g_A$ in medium and therefore set $g_A^* = 1.0$.

\begin{figure}[hbt]
\begin{center}
\includegraphics[height=15cm, angle=-90]{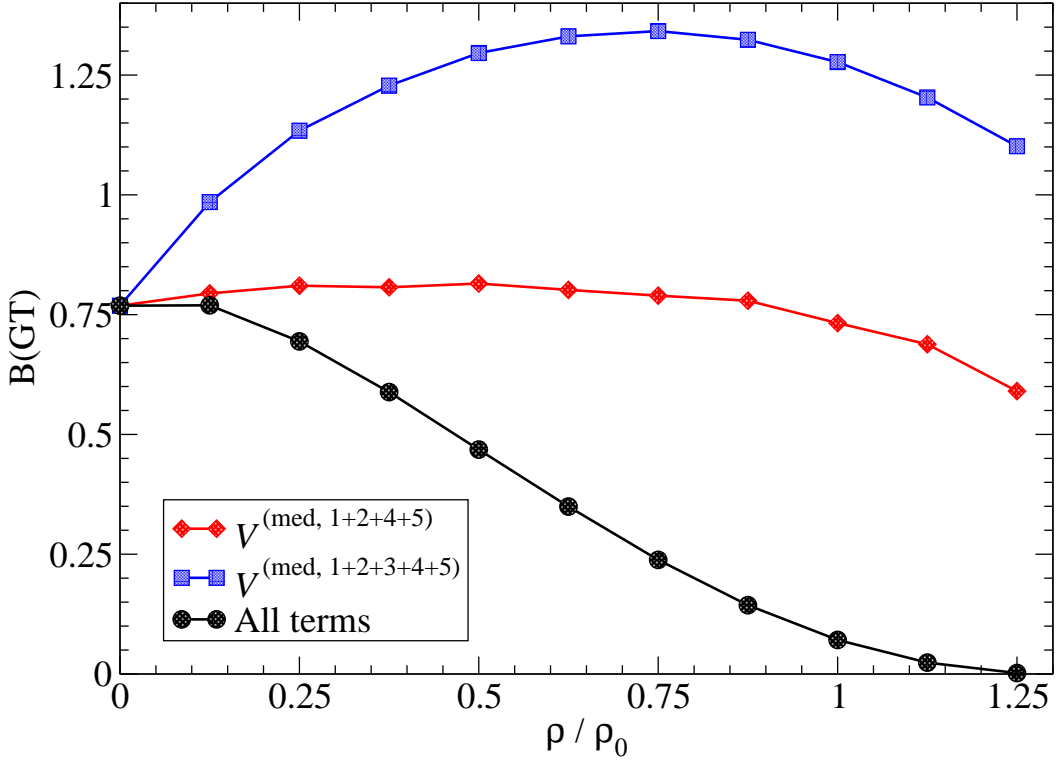}
\caption{The effects of the various density-dependent contributions to the in-medium NN interaction on the ground state to ground state $B(GT)$ value. The legend denotes which of the $V^{\rm med,i}_{NN}$ are included {\it together} with \vlkn.}
\label{bgtmed}
\end{center}
\end{figure}

In Fig.\ \ref{bgtmed} we separate the effects of the six different components of $V^{\rm med}_{NN}$ on the GT strength. Including only the sum of $V^{\rm med, 1}_{NN}, V^{\rm med, 2}_{NN}, V^{\rm med, 4}_{NN}$, and $V^{\rm med, 5}_{NN}$ together with the low-momentum interaction $V_{\rm low-k}$, there is very little density dependence in the calculated GT strength. These four density-dependent components are exactly the terms that are expected to approximately cancel at leading order in the effective shell model interaction. Including the additional $V^{\rm med, 3}_{NN}$, we find that there is a mild increase in the GT strength. This is in accordance with our predictions based on the first-order perturbative calculation of the $^{14}$C and $^{14}$N ground state wavefunctions shown in eq.(\ref{pert3}). However, given the very small mixing of the excited state wavefunctions at first-order, the second-order diagrams contributing to the shell model effective interaction are equally important. Finally, by including $V^{\rm med,6}_{NN}$ in addition to all other terms, we find that the GT strength is strongly suppressed. From eq.(\ref{pert6}) our interpretation is that the 3N contact interaction strongly shifts strength to the $\left | 33 \right >$ components of both the $^{14}$C and $^{14}$N ground state wavefunctions, thereby leading to the observed suppression.

\begin{figure}[hbt]
\begin{center}
\includegraphics[height=15cm, angle=-90]{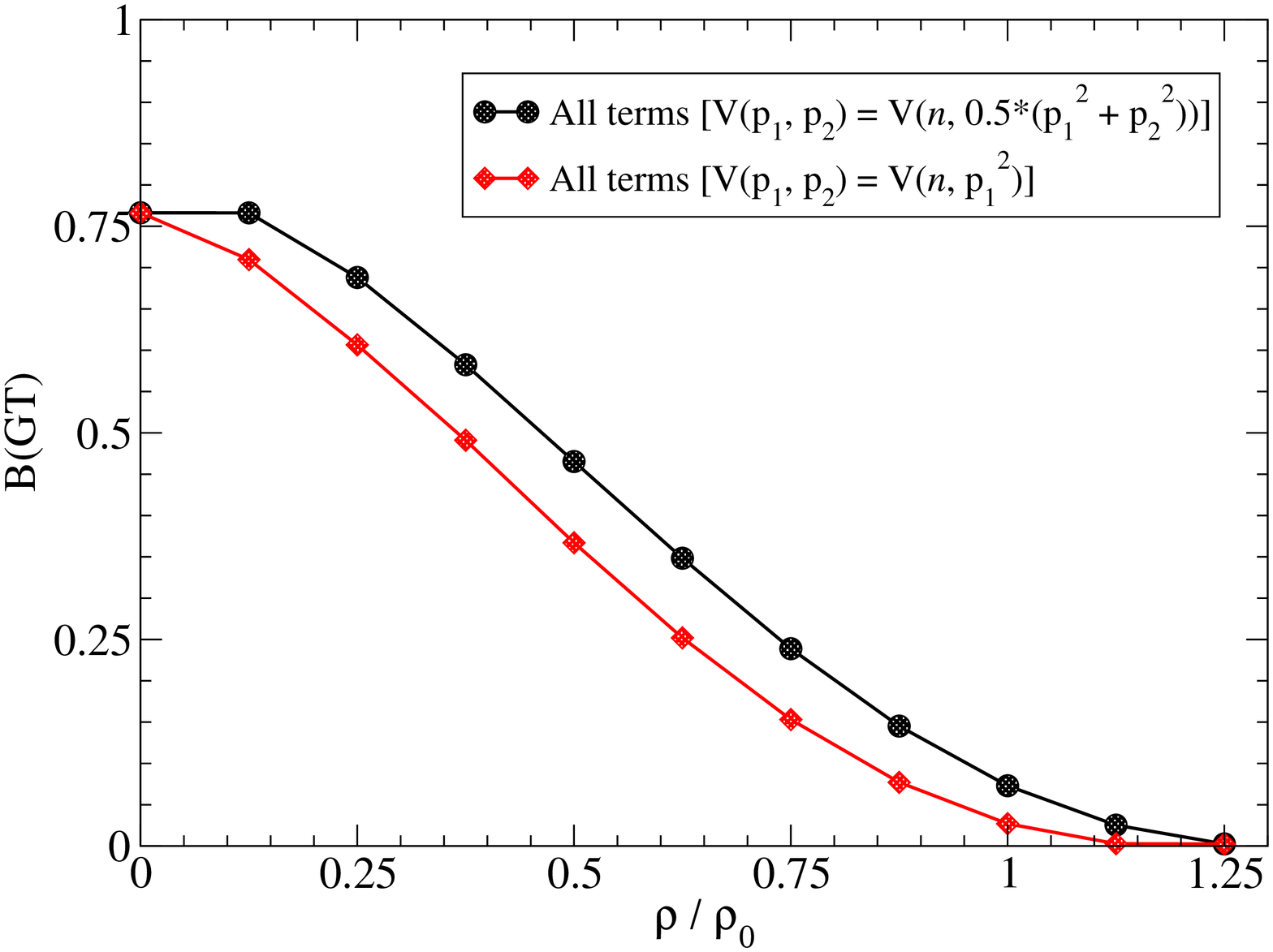}
\caption{Dependence of the ground state to ground state $B(GT)$ value on the off-shell extrapolation of the density-dependent $V^{\rm med}_{NN}$ interaction. Both the symmetric and asymmetric off-shell extrapolation are shown.}
\label{bgtoff}
\end{center}
\end{figure}
In Section 2 we suggested that the symmetric off-shell extrapolation of the density-dependent interaction is preferable to the asymmetric extrapolation, given the former's close agreement with the off-shell behavior of the OPE $V_{\rm low-k}$. In Fig.\ \ref{bgtoff} we study the effect of this choice on the Gamow-Teller strength. Regardless of the off-shell extrapolation method, we find a clear suppression of the $B(GT)$ value for densities close to that of nuclear matter. Indeed, the most important component of the density-dependent NN interaction, $V^{\rm med, 6}_{NN}$, is a contact interaction and is therefore independent of the off-shell extrapolation used. We conclude that the error associated with the off-shell prescription is rather small, given that even the extreme asymmetric extrapolation is rather close to the results from the symmetric extrapolation.

\begin{figure}[hbt]
\begin{center}
\includegraphics[height=15cm, angle=-90]{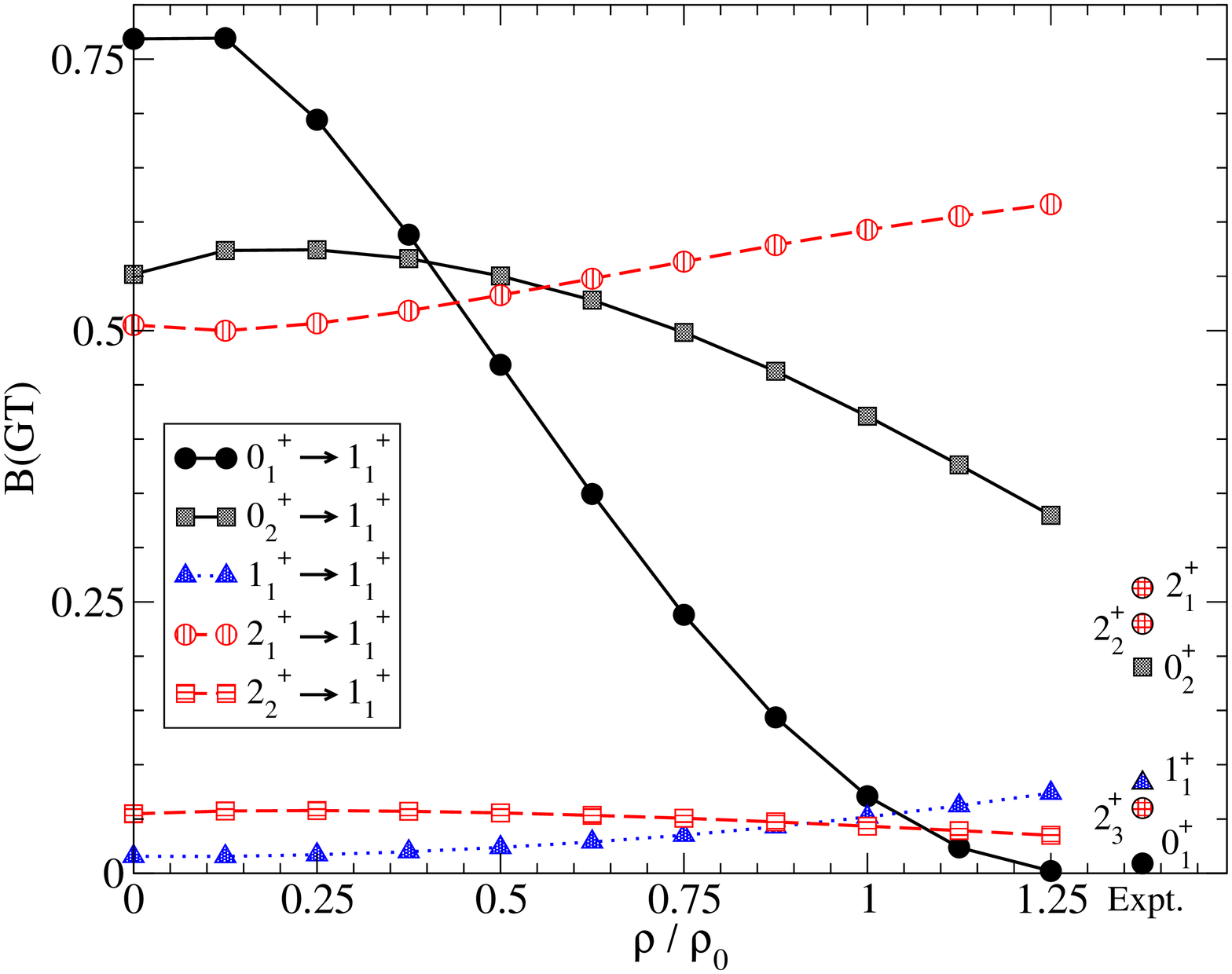}
\caption{The $B(GT)$ values for transitions from the low-lying states of $^{14}$C to the ground state of $^{14}$N
 as a function of the nuclear density. The
  experimental values are from \cite{negret}. Note that there are three
  experimental low lying $2^+$ states compared to two theoretical $2^+$ states
  in the $0p^{-2}$ configuration.}
\label{bgt}
\end{center}
\end{figure}

Given the strong dependence of the $^{14}$C beta decay transition rate on the nuclear density, it is important to study how other Gamow-Teller strengths are effected by the inclusion of $V_{NN}^{\rm med}$. Recently, GT strengths from the $^{14}$N ground state to excited states of $^{14}$C have been determined from the experimental charge exchange reaction $^{14}{\rm N}(d, {^{2}{\rm He}})^{14}{\rm C}$ \cite{negret}.
In Fig.\ \ref{bgt} we plot our results for the calculated $B(GT)$ values as a function of the nuclear density $\rho$. From Fig.\ \ref{bgt} one sees that the medium effects improve the agreement between our theoretical calculations and the experimental values for transitions from $0^+$ and $1^+$ states. The largest effect is clearly a suppression of the ground state to ground state transition for densities at and above that of nuclear matter. The other transition strengths are much less sensitive to the density dependence of the nuclear interaction.

\begin{figure}[hbt]
\begin{center}
\includegraphics[height=15cm, angle=-90]{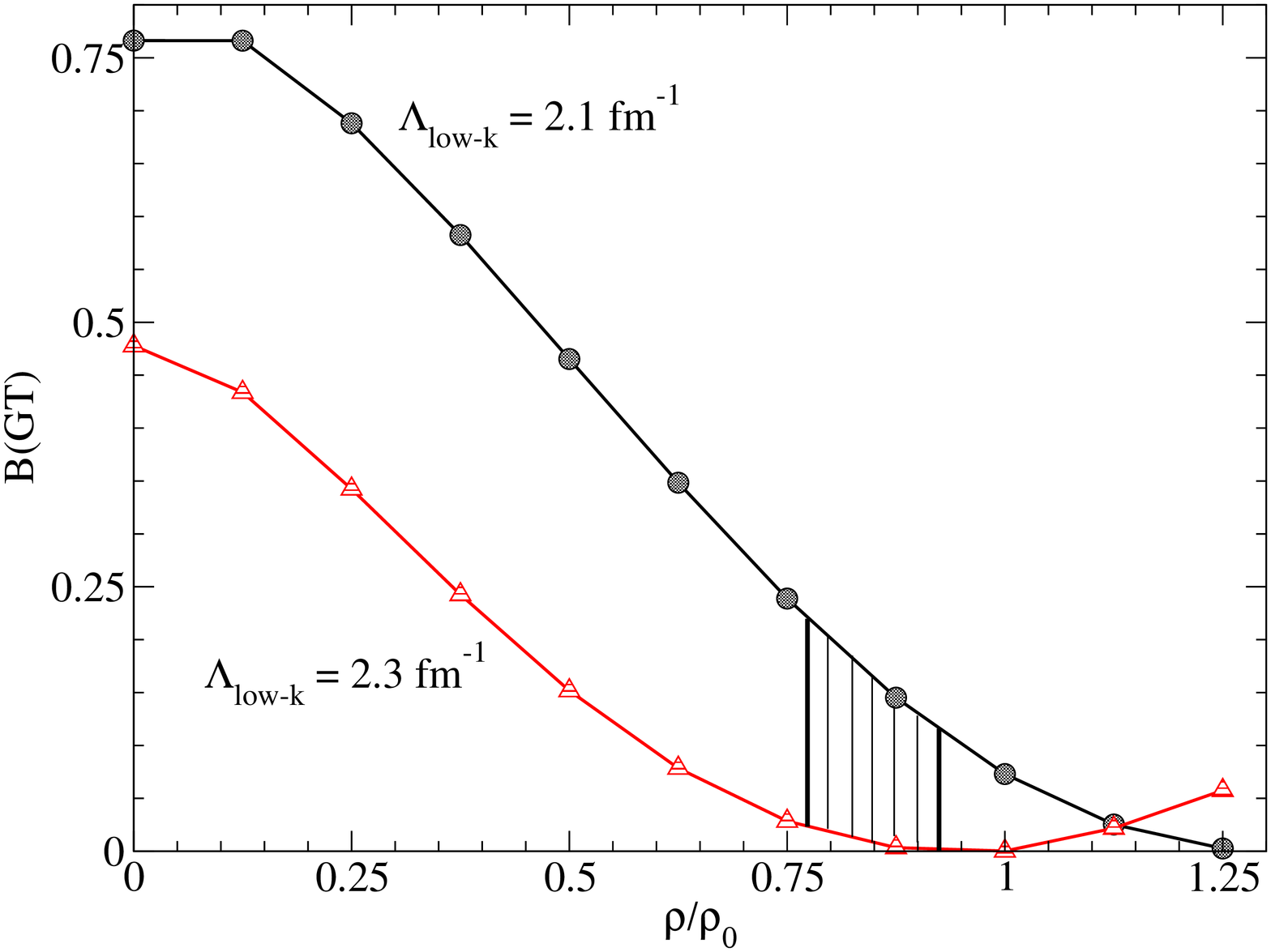}
\caption{The uncertainty in the calculated value of $B(GT)$ obtained by varying the low momentum cutoff $\Lambda_{\rm low-k}$ between 2.1 and 2.3 fm$^{-1}$. The shaded region corresponds to nuclear densities close to that experienced by valence $0p$-shell nucleons in $^{14}$C.}
\label{bgterror}
\end{center}
\end{figure}

Although we have chosen a low-momentum cutoff scale of $\Lambda_{\rm low-k} = 
2.1$ fm$^{-1}$ in the above calculations, the value of $c_E$ (and consequently 
the strength of the important $V^{\rm med, 6}_{NN}$ term) depends sensitively 
on $\Lambda_{\rm low-k}$ as seen in Table \ref{cdcelowk}. We therefore
calculate the beta decay transition strength using also a cutoff scale of $\Lambda_{\rm low-k} =2.3$ fm$^{-1}$, which incorporates some components of the bare NN interaction unconstrained by elastic NN scattering data but which provides an estimate for the error associated with the chosen momentum cutoff. The results for densities up to $1.25\rho_0$ are shown in Fig.\ \ref{bgterror} together with the calculations using $\Lambda_{\rm low-k} = 2.1$ fm$^{-1}$. We suggest that this constitutes the most important source of error in our calculation.  Additional errors include the following: deviations of the effective axial coupling constant $g_A^*$ from 1.0, the choice of the off-shell extrapolation for the in-medium NN scattering amplitudes, and finally, errors associated with our use of the shell model at second-order in perturbation theory. In earlier sections of the paper we have estimated errors associated with the first two approximations, but the last of the three is difficult to quantify. Nevertheless, we expect such errors to be minor, given that previous shell model calculations (see \cite{corag} and references therein) using \vlk up to second order in perturbation theory have been able to describe very well the properties of nuclei with a small number of valence nucleons above a closed shell.

\begin{figure}[hbt]
\begin{center}
\includegraphics[height=15cm,angle=-90]{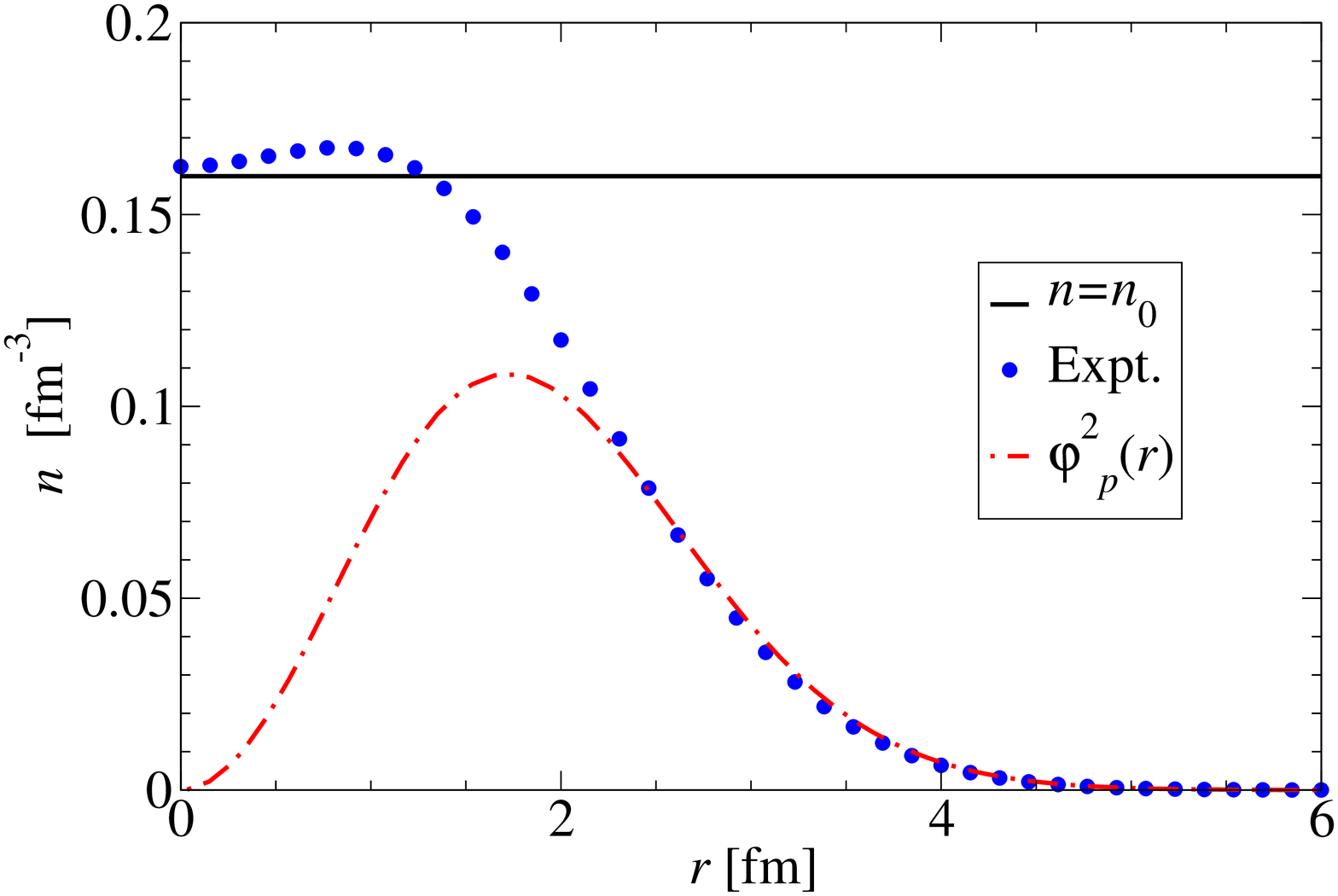}
\caption{Twice the charge distribution of $^{14}$N taken from \cite{schaller, schutz} together with the square of the radial $0p$-shell wavefunctions.}
\label{hosc}
\end{center}
\end{figure}

Since the theoretical value of $B(GT)$ is particularly sensitive to the nuclear density, let us estimate the range over which we expect our calculations to be valid. In Fig.\ \ref{hosc} we plot twice the charge distribution of $^{14}$N
obtained from electron scattering experiments \cite{schaller,schutz} fit
to the harmonic oscillator density distribution
\be
n(r) \propto \left(1 + b\frac{r^2}{d^2} \right) e^{-r^2/d^2},
\ee
where for $^{14}$N, $b = 1.291$ and $d=1.740$ fm.
We compare this density distribution to the square of the $0p$-shell radial wavefunction used in our calculation. This wavefunction peaks at a nuclear density of approximately $\rho = 0.85\rho_0$. We have therefore shaded the region of Fig.\ \ref{bgterror} corresponding to this density $\pm$ approximately 10\%.

\section{Summary and outlook}
We have studied the effects of the leading-order chiral three-nucleon force on the beta decay lifetime of $^{14}$C, and more generally on the Gamow-Teller transition strengths from the ground state of $^{14}$N to the low-lying states of $^{14}$C. Our results indicate that the density-dependent in-medium NN interaction $V^{\rm med}_{NN}$ derived from the chiral 3N force has a strong effect on the ground state to ground state transition. In contrast, the GT strengths from the ground state of $^{14}$N to the excited states of $^{14}$C exhibit only a small density dependence. These results are consistent with the calculations presented in \cite{holt}.
We find that the GT strength is particularly sensitive to the genuine short-range component of the chiral 3NF (which is almost completely responsible for driving the suppression) as well as the low-momentum decimation scale $\Lambda_{\rm low-k}$. In general, by fitting the binding energies of $A=3,4$ nuclei, one can derive a constraint curve relating the parameters $c_D$ and $c_E$ \cite{nav}. Additional information is then required to fix the point on this curve. Given the sensitivity of the $^{14}$C lifetime on the parameter $c_E$, we suggest that this decay can serve as a useful constraint on the two low-energy constants $c_D$ and $c_E$ once {\it ab initio} many-body calculations are able to study this problem more accurately.

\end{document}